\documentclass[prl,twocolumn,superscriptaddress,preprintnumbers,amsmath,amssymb]{revtex4}
\usepackage{graphicx}

\begin{document}
\title{Electron Correlations and the Minority-Spin Band Gap in Half-Metallic Heusler Alloys}
\author{L.~Chioncel}
\affiliation{Institute of Theoretical Physics, Graz University of Technology, A-8010 Graz Austria}
\author{E.~Arrigoni}
\affiliation{Institute of Theoretical Physics, Graz University of Technology, A-8010 Graz Austria}
\author{M.I.~Katsnelson}
\affiliation{Radboud University Nijmegen, NL-6525 ED Nijmegen, The
Netherlands}
\author{A.I.~Lichtenstein}
\affiliation{Institute of Theoretical Physics, University of Hamburg, Germany}
%\pacs{71.15.Ap;71.10.-w;73.21.Ac;75.50.Cc}

\begin{abstract}
Electron-electron correlations affect the band gap of
half-metallic ferromagnets by introducing non-quasiparticle states
just above the Fermi level. In contrast to the spin-orbit
coupling, a large asymmetric non-quasiparticle spectral weight is present in
the minority-spin channel, leading to a peculiar
finite-temperature spin depolarization effects. Using recently
developed first-principle dynamical mean field theory, we
investigate these effects for the half-metallic ferrimagnetic
Heusler compound FeMnSb. We discuss depolarization effects in
terms of strength of local Coulomb interaction $U$ and temperature
in FeMnSb. We propose Ni$_{1-x}$Fe$_{x}$MnSb alloys 
as a perspective materials to be used in spin-valve structures and for 
experimental search of non-quasiparticle states in half-metallic materials.
\end{abstract}
\maketitle

The realization of physical systems whose electronic states can be
easily manipulated by external actions is a very promising issue
both from fundamental and applied side. For example, spintronics
materials in which charge and spin degrees of freedom can be used
simultaneously in order to produce devices with new functionality,
have become subjects of growing interest.

A family of extensively studied spintronics materials are the
half-metallic compounds (HM). First principle electronic structure 
calculations for HM showed 
unusual properties in their spin-resolved band structure: while the
electronic states for one spin projection have a metallic
character with a nonzero density of states at the Fermi level
$E_F$, the states with the other spin projection demonstrate a
band gap around $E_F$%, just like in a semiconductor or insulator
\cite{deGroot83,ufn}. As a result, HM can in principle conduct a
fully spin-polarized current, and therefore attract much attention
due to potential applications in the field of spintronics
\cite{ufn,sarma1}. These calculations, based on the density-functional 
theory whithin the Local Density Approximation (LDA) or
the Generalized Gradient Approximation (GGA),
are very successful in many cases to describe or predict material
properties. However, they fail notably for the case of strongly-correlated
electron systems. For such
systems the LDA+DMFT (Dynamical Mean-Field Theory) method has been
designed \cite{anisDMFT,ourDMFT,Katsnelson99} and currently is
used very extensively for various applications \cite{today}.

One of the dynamical many-electron features of HMF, the
non-quasiparticle states \cite{ufn,edwards,IK}, contribute
significantly to the tunneling transport in heterostructures
containing HMF \cite{ourtransport,falko}, even in the presence 
of arbitrary disorder \cite{surface}. The origin of these states 
is connected with ``spin-polaron'' processes: the spin-down
low-energy electron excitations, which are forbidden for HMF in
the one-particle picture, turn out to be possible as
superpositions of spin-up electron excitations and virtual magnons
\cite{edwards,IK}. In  previous publications, we applied the LDA+DMFT approach to
describe the non-quasiparticle states in NiMnSb \cite{Chioncel03} 
and CrAs \cite{Chioncel05} and discussed the possible 
experimental investigation (Bremsstrahlung isohromat spectroscopy, 
nuclear magnetic resonance, scanning tunneling microscopy, and 
Andreev reflection) techniques to clarify the existence of these states
\cite{Chioncel03}.

Here, we investigate the distinction between static (i.e.
spin-orbit) and dynamic (correlation) effects leading to a finite
temperature depolarization. According to our results, for the class 
of Heusler HM, the strong depolarization at finite temperatures is 
essentially due to correlation effects, while the spin-orbit interaction 
gives a negligible effect. In addition, we identify characteristic 
features of the DOS spectrum that should help distinguishing between 
static and dynamic depolarization effects.

Crystal imperfections \cite{Ebert91}, interfaces \cite{Wijs01},
and surfaces \cite{Galanakis03} constitute important examples of
static perturbations of the ideal, periodic potential which affect
the states in the half-metallic gap. It was shown recently
\cite{spirals}, that due to finite temperatures, the static
non-collinear spin-configurations shows a mixture of spin-up and
spin-down density of states that destroy the half-metallic
behavior. The non-collinearity could exist at zero temperature as
well, in anisotropic structures, due to the spin-orbit coupling.
In such cases static spin-flip scattering will introduce states in
the half-metallic gap. It is the purpose of the present paper to use 
a many-body approach to investigate dynamical spin fluctuation effects on the
electronic structure at temperatures below the Curie temperature,
$T<T_c$, within the half-metallic state.
In order to illustrate the specific differences between the many-body
and the static non-collinear effects we extend here our previous
LDA+DMFT calculations \cite{Chioncel03,Chioncel05} to a different
half-metallic Heusler alloy, that is, ferrimagnetic FeMnSb. One of
the motivations to study the hypothetical FeMnSb material is to
explore the effects of Fe doping in the host NiMnSb compound.

In our approach, correlation effects are treated in the
framework of dynamical mean field theory (DMFT) \cite{today}, with
a  spin-polarized T-matrix Fluctuation Exchange (SPTF) type of the
DMFT solver \cite{Katsnelson99}, the on-site Coulomb interaction
being described by full four-indices orbital matrix \cite{ourDMFT}.
The essential quantity is the LDA+DMFT bath Green function ${\cal
G}^{\sigma}_{0,m,m'}$, where $0$ denotes the impurity site. The
DMFT self-consistency equation ${\cal
G}^{-1}_{0,mm'}={G}^{-1}_{LDA,mm'}-{\Sigma}_{0,mm'}$ is used to
combine the LDA Green function with the solution of the impurity
model ${\Sigma}_{0,mm'}={\Sigma}_{0,mm'}[{\cal G}_{0,mm'}]$.
Further computational details are described in Refs.
\onlinecite{Chioncel03,Chioncel05,EMTODMFT}.

In the simplest case of neglecting the dispersion of the magnon
frequency, $\omega _{\mathbf{q}}\approx \omega _{m}$, in comparison
with the electron hopping energy $t_{\mathbf{k}}$, the electronic
self-energy becomes local \cite{ufn,IK}: For such a self-energy
one can calculate the one-electron Green functions
$G_{\mathbf{k},\downarrow }(E)=[E-t_{\mathbf{k},\downarrow }-
\Sigma _{\downarrow }^{loc}(E)]^{-1}$, which allow us to evaluate
the additional contributions to the density of states. For illustrative 
purposes we write the lowest-order contribution:
\begin{equation}\label{nnqp}
\delta N_{\downarrow}(E) = \sum_{\mathbf{k}} Re \Sigma_{\downarrow}(E)
\delta^\prime(E-t_{\mathbf{k}\downarrow}) - \frac 1 \pi
\sum_{\mathbf{k}} \frac{Im \Sigma_\downarrow(E)}{(E-t_{\mathbf{k}\downarrow})^2}
\end{equation}
The second term of Eq.(\ref{nnqp}) is formally connected with the
branch-cut of the Green function due to the electron-magnon
scattering processes \cite{ufn,IK}.

In general, one should take into account spin-orbit coupling effects 
connecting the spin-up and -down channels through the
scalar product between the angular momentum ${\bf l}$ and the spin
${\bf s}$ operators. The strength of this interaction is proportional to the spatial
derivatives of the crystal potential $V({\bf r})$: $V_{SO} \propto grad V~({\bf l}\cdot{\bf s})$
with non zero off-diagonal elements ${V}^{\sigma,\sigma^\prime }$, $\sigma=\uparrow, \downarrow$. 
For HM with a gap in the minority-spin (spin down) channel, one could
construct the wave function for spin-down electrons based on the
general perturbation consideration so that the density of states (DOS) 
in the gap has a quadratic dependence of the spin orbit coupling strength
\cite{Mav04}:
%\begin{equation}\label{nso}
$\delta n^{SO}_{\downarrow}(E) \propto ({V}^{\downarrow,\uparrow})^2$
%\end{equation}
As one can see there is an obvious qualitative 
distinctions between the many-body (Eq.\ref{nnqp}) and the spin-orbit contribution
in the minority spin channel, in addition to the fact
that the latter are orders in magnitude smaller
\cite{Mav04}. In the former case the strong temperature dependence
of the residues of the Green function and the ``tail'' of the NQP
states give rise to a strong temperature dependence of the spin
polarization, while the spin-orbit term is very weakly temperature 
dependent.

As a matter of fact, in the non-relativistic approximation (without
spin-orbit coupling) there are two essentially different sources
for states in the gap at finite temperatures. First, there is
the simple classical effect of band filling due to disorder, that
is, due to scattering on static (classical) spin fluctuations.
This kind of the gap filling is symmetric with respect to the
Fermi energy, that is, there is no essential difference between
electrons and holes. Contrary, the correlation effects
result in an asymmetry in the gap filling \cite{ufn}, namely, NQP
states appear in spin-down channel just above the Fermi level
\cite{edwards,ufn,IK,Chioncel03,Chioncel05}, and in the spin-up
channel below the Fermi level \cite{ufn,IK}. This asymmetry is
a purely quantum effect connected with the Pauli principle and
with the quantum character of spins; it disappears in the classical limit
\cite{ufn}. For example, in the case under consideration
(minority-spin gap) from a consequent quantum
point of view, the conduction-electron spin projection is not a
good quantum number and real electron states are superpositions of
the minority-electron states and majority-electron states plus
magnon, with the same projection of the {\it total} spin of the
crystal. However, the majority-electron states {\it below} the
Fermi energy cannot participate in this superposition since they
are already completely occupied, therefore these quantum
(spin-polaronic) effects below the Fermi energy is totally
suppressed at zero temperature by the Pauli principle. As a
result, the states in the minority-spin gap at finite temperatures
are formed from these spin-polaron states existing only above the
Fermi energy plus the disorder-smeared states filling the gap more
or less uniformly.

\begin{figure}[h]
\includegraphics[angle=270,width=\linewidth]{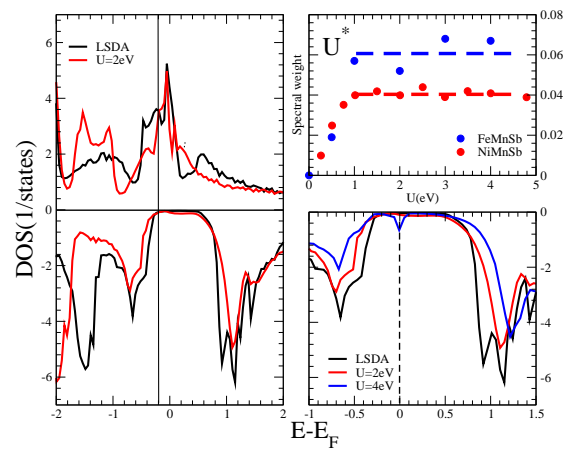}
\caption{Left: Density of states of half-metallic {\it FeMnSb},
LSDA (black line) and LSDA+DMFT (red line), for the efective Coulomb
interaction $U=2$eV exchange parameter $J=0.9$eV and temperature
$T=300K$. Lower right panel: zoom around $E_F$ for different values
of $U$. Upper right panel: Spectral weight of the NQP states calculated
as function of $U$. The values obtained for {\it NiMnSb}
\cite{Chioncel03} are plotted for comparison.
\label{fig:DOSFeMnSb}}
\end{figure}

Early theoretical studies %on the half-metallic Heusler alloys
showed that the %electronic structure and the 
gap of the minority
spin channel is stable with respect to differently chosen {\it 3d}
atom X=Fe,Co,Ni in the XMnSb compounds \cite{deGroot86,Kubler84}.
Notable difference between Ni- and Fe-based Heusler alloys is that
NiMnSb is a ferromagnetic half-metal, with a very small value of
Ni magnetic moment ($0.2\mu_B$), whereas in FeMnSb the
ferrimagnetic coupling between Fe ($-1\mu_B$) and Mn ($3\mu_B$)
moments stabilizes the gap and the half-metallic electronic
structure% in the spin-down channel 
\cite{deGroot86}.

In the calculations we considered the standard representation of
the $C1_b$ structure with a fcc unit cell containing three atoms:
Fe$(0,0,0)$, Mn$(1/4,1/4,1/4)$, Sb$(3/4,3/4,3/4)$ and a vacant
site E$(1/2,1/2,1/2)$, respectively. Unfortunately the ternary
compound FeMnSb does not exist, however indications concerning
magnetic and crystallographic properties were obtained by
extrapolating the series of Ni$_{1-x}$Fe$_{x}$MnSb
\cite{deGroot86}, to high Fe concentration. In this case we chose
a lattice parameter of $a=5.882 \AA$ for FeMnSb the same as in the
recent LDA+SO calculation of Mavropoulos {\it et.al.}
\cite{Mav04}. To illustrate the differences between the static and
dynamic effects we plot the DOS of the LDA+DMFT calculations which
should be compared with recent results including SO coupling
\cite{deGroot86,Mav04}.

Note that depending on the character of chemical bonding, the 
value of $U$ for all 3d metals is predicted to vary 
between 2 and 6-7 eV \cite{today}
A relatively weak dependence of the non-quasiparticle spectral
weight, on the $U$ value, (Fig. \ref{fig:DOSFeMnSb}) is evidenced
for both NiMnSb and FeMnSb compounds. A ``saturation'' of the
spectral weight of FeMnSb takes place for almost the same value,
$U^* \approx 1eV$ as in the case of NiMnSb. This effect is
understood in terms of the $T$-matrix renormalization of the
Coulomb interactions \cite{Katsnelson99}. The
spectral weight values for FeMnSb are larger in comparison with
the ones obtained for NiMnSb \cite{Chioncel03}, which can be
attributed to the fact that at the Fermi level a larger DOS is
present in the ferrimagnetic FeMnSb than in the ferromagnetic
NiMnSb. 

The spin-orbit coupling produce a peak close to the Fermi level
\cite{Mav04} in the minority-spin channel, which is an order of
magnitude smaller than the spectral weight of the NQP states.
According to the SO results \cite{Mav04}, the polarization at the 
Fermi level for NiMnSb and FeMnSb are almost the same. In contrast, 
our calculation shows that the spectral weight of NQP states
in FeMnSb is almost twice as large as the value calculated for
NiMnSb. 

\begin{figure}[h]
\includegraphics[angle=0,width=0.95\linewidth]{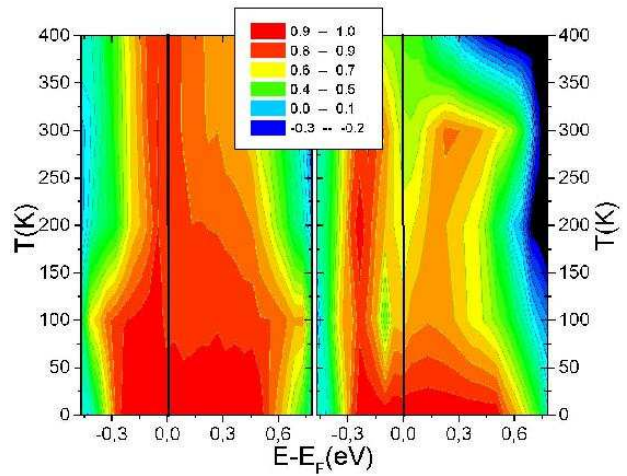}
\caption{Contour plots of polarization as function of energy and temperature
for different values of local Coulomb interaction U. Left $U=2eV$, right
$U=4eV$. The LSDA polarization is plotted as the $T=0K$ temperature result. The asymmetry
of the NQP states, is clearly visible for $U=4eV$.
\label{pol_t}}
\end{figure}

In order to discuss the influence of temperature and local Coulomb
interactions, on the polarization in FeMnSb compound, we present
results of LDA+DMFT calculations for $T \leq 400K$, and different
$U$'s. Fig \ref{pol_t} presents the contour plot of polarization
$P(E)=(N_{\uparrow}(E)-N_{\downarrow}(E))/(N_{\uparrow}(E)+N_{\downarrow}(E))$
as a function of temperature $T$ for $U=2$ and $4eV$. The LDA value, 
plotted for convenience as the $T=0$ result shows %the presence of 
a gap of magnitude $0.8eV$ in agreement with previous calculations 
\cite{Mav04}.

One can see a peculiar temperature dependence of the spin
polarization. The NQP features appears for $E-E_F \geq 0$, 
and is visible in Fig. \ref{fig:DOSFeMnSb}, for $U=2eV$ and 
the temperature $T=300K$. A strong depolarization effect is 
evidenced for the larger value of $U=4eV$. Already at 100K, 
there is a strong depolarization of about 25$\%$. Increasing 
the value of $U$, form $2eV$ to $4eV$, the non-quasiparticle 
contribution is more significant, therefore the
predominant factor in depolarization is played by the NQP. When
the tail of these states crosses the Fermi level a drastic
depolarization at Fermi level takes place. One can notice
that for the case of $U=4eV$, the NQP is pinned at the Fermi
level, and has a large contribution also due to a large value of
DOS in the spin up channel.

\begin{figure}[h]
\includegraphics[angle=0,width=0.95\linewidth]{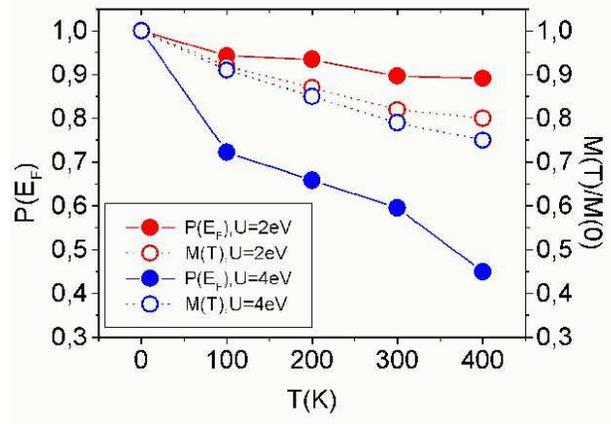}
\caption{Temperature dependent polarization at the Fermi level,
$P(E=E_F, T)$ (solid line) and magnetization (dashed
line) for different values of local Coulomb interaction U.
\label{pm_t}}
\end{figure}

In Fig. \ref{pm_t} one can see a clear distinction between the
finite-temperature behavior of the polarization and magnetization, 
for different values of $U$. It is interesting to note that the 
reduced magnetization $M(T)/M(0)$ decreases slowly  in the temperature 
range studied in Fig. \ref{pol_t}. This reduction is a consequence of 
the finite temperature excitations, i.e spin-flip processes, affecting 
both spin channels. In the minority spin channel, NQP states are formed, 
and in the majority channel a spectral weight redistribution around 
the Fermi level (Fig.\ref{fig:DOSFeMnSb}) contributes to the depolarization. 
The corresponding depolarisation increases with the strength of correlations.
The density of NQP states displays a rather strong temperature dependence 
\cite{IK,ourtransport}, resulting in the asymmetry that is visible in 
Figs.~\ref{fig:DOSFeMnSb} and \ref{pol_t}.  Recently Dowben et. al. \cite{spirals} 
showed that non-collinearity results in a spin mixing which ultimately
leads to a nonvanishing but symmetric DOS around the Fermi level  in the 
gap of the insulating spin channel. Therefore, we suggest that the asymmetry 
in the DOS of the minority channel is the key feature that should help 
distinguishing whether correlation effects are responsible for the 
finite-temperature depolarisation or not.  

Our results show that at finite temperatures the NQP states appear
in the gap of the minority spin channel, reducing the polarization
significantly. In such a situation,  the half-metallic state is depleted.
On the other hand, recently, Wang \cite{Wang99} suggested that states
in the minority-spin gap of a half-metallic ferromagnet are localized 
due to disorder (Anderson localization). In such case, the material 
behaves like a fully polarized HFM. The question whether  NQP states 
are conducting or not has not been studied in the present work, and 
will be the subject of our future research.

Our work suggests that  depolarisation in this class of Heusler
compounds  is dominated by NQP states, while  spin-orbit 
contributions are much smaller. In addition, many-body effects are 
more pronounced in FeMnSb than in NiMnSb. This is tightly connected 
to the larger DOS in the majority spin channel in the former material.
Therefore, doping of NiMnSb by Fe could be an interesting issue to 
investigate the interplay between alloying and many body effects. 
In this respect, we have carried out preliminary LDA+DMFT 
calculations \cite{private} on NiMnSb supercell containing 25 \% Fe 
impurities, i. e. for  (Ni$_3$Fe)Mn$_4$Sb$_4$. Our results show a 
half-metallic character at the  LDA level, with similar strong 
correlation-induced depolarization effects as in pure FeMnSb. Therefore, 
for this material, many body effects are of primary importance even in 
the presence of disorder. Correlation effects on surfaces of half-metals 
were dicussed recenty and it was shown that these states can be probed 
both directly and via their effect on surface states \cite{surface}.
As a consequence in addition to the previously discussed 
experimental techniques \cite{Chioncel03}, we propose the use of 
Ni$_{1-x}$Fe$_{x}$MnSb alloys both in spin-valve 
structures, and to investigate the existence of NQP states 
in half-mettalic materials. 

The discovery of giant magneto-resistance ({\it GMR}) has led to
tremendous amount of activity to understand and develop technology 
based on high-density magnetic recording. The dominant mechanism 
leading to GMR is the spin-dependent {\it s-d} scattering. NiMnSb-based 
spin-valve structures using Mo spacer layers NiMnSb/Mo/NiMnSb/SmCo$_2$ were
successfully produced \cite{Hordequin98}. The associated {\it GMR}
exhibits a clear spin-valve contribution of around $\Delta R/R
\approx 1\%$ \cite{Hordequin98}. One of the limiting factor for
such a small value is the large resistivity of the Mo layer which
determines limited flow of {\it active electrons} exchanged
between the two ferromagnetic layers without being scattered. To
improve on the {\it GMR} value the use a low-resistivity
standard spacer such as {\it Au} or {\it Cu} was suggested \cite{Hordequin98}. 
On the other hand, having a larger value of DOS at the Fermi level which 
occurs in the majority spin channel (Fig. \ref{fig:DOSFeMnSb}), the ferrimagnetic
FeMnSb or Ni$_{1-x}$Fe$_x$MnSb could increase the number of {\it active electrons} in such
a spin valve configuration. In addition, due to its ferrimagnetic
properties, the spin-valve demagnetization field can be reduced.
The reduced demagnetization is extensively exploited in synthetic
ferrimagnet spin valves heads and are known to have advantages
over conventional spin-valve heads \cite{Kanai01}.

It is interesting to note that, due to its larger DOS in the
majority spin channel, FeMnSb is expected on the one hand to provide 
a better performance in Half-metal based spin-valves in comparison 
with NiMnSb. On the other hand, our calculation shows that such a 
larger DOS is accompanied by an equally larger DOS of NQP that, on 
the other hand, suppresses polarisation. The conclusion, thus, is 
that correlation effects are important and should not be neglected 
precisely in putatively "good" HM materials, i. e. in materials with 
a large DOS in the majority spin channel. 

{\it Acknowledgement} We thank Dr. R.A. de Groot for helpful
discussions, and acknowledge Dr. Ph. Mavropoulos for providing 
his LDA+SO results. We acknowledge financial support by
the KFA Juelich (LC) and by the FWF project P18505-N16 (LC and EA).

\end{document}